\documentclass[seceq]{ptptex}
\usepackage{wrapft}

\newcommand{\beq}{\begin{equation}}
\newcommand{\eeq}{\end{equation}}
\newcommand{\beqa}{\begin{eqnarray}}
\newcommand{\eeqa}{\end{eqnarray}}
\newcommand{\bra}[1]{\mbox{$\langle #1|$}}
\newcommand{\ket}[1]{\mbox{$|#1\rangle$}}

\usepackage{graphicx}
\usepackage{psfig}
\usepackage{epsfig}

\pubinfo{Vol.~113, No.~1, January 2005}
\notypesetlogo  

\title{{\it NN} Final-State Interaction in the Helicity Dependence of
  Inclusive $\pi^-$ Photoproduction from the
  Deuteron\footnote{Dedicated to Prof.\ H.\ Arenh\"ovel on the
    occasion of his 66th birthday.}}

\author{Eed M.\ \textsc{Darwish}\footnote{{\rm E-mail address:}
    eeddarwish@yahoo.com}}

\inst{Physics Department, Faculty of Science, South Valley University, \\
  Sohag 82524, Egypt}

\recdate{October 19, 2004}

\abst{The helicity dependence of the inclusive $\pi^-$ photoproduction
  reaction from the deuteron in the $\Delta$(1232)-resonance region is
  investigated with inclusion of final-state nucleon-nucleon
  rescattering ($NN$-FSI). For the elementary $\pi$-production
  operator an effective Lagrangian model which includes the standard
  pseudovector Born terms and a contribution from the
  $\Delta$-resonance is used. The half-off-shell $NN$-scattering
  matrix is obtained from a separable representation of a realistic
  $NN$-interaction. The differential polarized cross-section
  difference for parallel and antiparallel helicity states is
  predicted and compared with experiment. We find that the effect of
  $NN$-FSI is much less important in the helicity difference than in
  the previously studied unpolarized differential cross section.
  Furthermore, the contribution of $\vec d(\vec\gamma,\pi^-)pp$ to the
  deuteron spin asymmetry is explicitly evaluated with inclusion of
  $NN$-FSI. It has been found that the effect of $NN$-FSI is much
  larger in the asymmetry than in the total cross section, and this
  leads to an appreciable reduction of the spin asymmetry in the
  $\Delta$-region. Inclusion of such effect also leads to improved and
  quite satisfactory
  agreement with existing experimental data.} 

\begin{document}

\maketitle

\section{Introduction}\label{sec1}
A very interesting topic in intermediate energy nuclear physics is
concerned with the quasi-free pion production reaction in nuclei,
which is governed by three main mechanisms: (i) the elementary
amplitudes of the four pion production channels possible from the
nucleon, (ii) the Fermi motion of the protons and neutrons inside the
nucleus, and (iii) the interaction between the final-state hadrons.
The investigation of pion photo- and electro-production has the
potential to become an important topic in meson physics, because many
important features of electromagnetic and hadronic reactions can be
studied through these processes. Interest in this topic has increased
mainly as a result of the construction of new high-duty continuous
electron beam machines such as MAMI in Mainz and ELSA in Bonn.

The particular interest in pion photo-production reaction from the
deuteron is due to the fact that the simple and well known deuteron
structure allows one to obtain information regarding the production
process from the neutron, which otherwise is difficult to obtain
because of the lack of free neutron targets. The earliest calculations
for pion photo-production from the deuteron were performed using the
impulse approximation (IA) \cite{ChL51,LaF52}. Approximate treatments
of final-state interaction (FSI) effects within a diagrammatic
approach have been reported in Refs.\ \cite{BlL77,Lag78,Lag81}. The
authors of those works noted that the FSI effects are quite small for
the charged-pion production channels in comparison to the neutral one.
Photo-production of pions from the deuteron has been investigated with
the spectator nucleon model \cite{ScA96}, ignoring all kinds of FSI
and two-body processes. The $NN$-FSI is considered in Ref.\ 
\cite{Lev01}, and good agreement with experiment was obtained. The
influence of final-state $NN$- and $\pi N$-rescattering on the
unpolarized cross sections is investigated in Ref.\ \cite{Dar03}.
There, it is found that $\pi N$-rescattering is much less important
(in general negligible) than $NN$-rescattering. Inclusion of such
effects leads to good agreement with experiment. The role of the
$N\Delta$-FSI in pion photo-production from the deuteron is
investigated in Ref.\ \cite{Faes02}. It has been shown that full
calculations with the off-shell amplitudes of $NN$- and $N\Delta$-FSI
are necessary to obtain a quantitative description of the cross
sections.

To this time, most of calculations have treated only unpolarized
observables, like the differential and total cross sections. These
cross sections provide information only regarding the sum of the
absolute squares of the amplitudes, whereas the polarization
observables allow extraction of more information. Observables with a
polarized photon beam and/or polarized deuteron target have not been
throughly investigated. The particular interest in these observables
is due to the fact that a series of measurements of the polarization
observables in photo-production reactions have already been carried
out and are planned at different laboratories. The GDH collaboration
has undertaken a joint effort to experimentally verify the
Gerasimov-Drell-Hearn (GDH) sum rule, measuring the difference between
the helicity components in the total and differential photo-absorption
cross sections. Our goal is to carry out an analysis of these
experimental measurements.

Recently, polarization observables for incoherent pion
photo-production reaction from the deuteron have been studied in
Refs.\ \cite{Log00,Dar03+,Dar04,Dar04E13,Dar04JG,Log04}. The
$\pi^-$-production channel has been studied within a diagrammatic
approach \cite{Log00} including $NN$- and $\pi N$-rescattering. In
that work, predictions for the analyzing powers connected to beam and
target polarization and to the polarization of one of the final
protons are presented. In a previous evaluation \cite{Dar03+}, special
emphasis was given to the beam-target spin asymmetry and the GDH sum
rule. Single- and double-spin asymmetries for incoherent pion
photo-production reaction from the deuteron are predicted in Refs.\ 
\cite{Dar04,Dar04E13,Dar04JG} without any FSI effects. The target
tensor analyzing powers of the $d(\gamma,\pi^-)pp$ reaction have been
studied in the plane wave impulse approximation \cite{Log04}. Most
recently, our evaluation \cite{Dar03+} was extended to higher energies
in Ref.\ \cite{Aren04} with additional inclusion of two-pion and eta
production.

As a further step in this study, we investigate in this paper the
influence of $NN$ FSI effects on the polarized differential and total
cross sections with respect to parallel and antiparallel spins of the
photon and the deuteron in the reaction $\gamma d\to\pi^-pp$. Our
second point of interest is to analyze the recent experimental data
from the GDH collaboration \cite{Pedroni}. With respect to the
interactions in the final two-body subsystems, only the
$NN$-rescattering is taken into account, because $\pi N$-rescattering
is considered negligible \cite{Lev01,Dar03}.

In \S\ref{sec2}, the model for the elementary $\gamma N\to\pi N$ and
$NN\to NN$ reactions that serves as an input for the reaction from the
deuteron is briefly reviewed. In \S\ref{sec3}, we introduce the
general formalism for incoherent pion photo-production from the
deuteron. The separate contributions of the IA and the
$NN$-rescattering to the transition matrix are described in that
section. Details of the actual calculation and the results are
presented and discussed in \S\ref{sec5}.  Finally, a summary and
conclusions are given in \S\ref{sec6}.
\section{The elementary $\gamma N\to\pi N$ and $NN\to NN$
reactions}\label{sec2} 
Pion photo-production reaction from the deuteron is governed by basic
two-body processes, namely pion photo-production from a nucleon and
hadronic two-body scattering reactions. For the latter, only
nucleon-nucleon scattering is considered in this work. As mentioned in
the introduction, $\pi N$-rescattering is found to be negligible, and
therefore it is not considered in the present calculation.

The starting point for the construction of an operator for pion
photo-production in the two-nucleon space is the elementary pion
photo-production operator acting on a single nucleon, i.e.\ $\gamma
N\to\pi N$. In the present work we examine various observables for
pion photo-production reaction from the free nucleon using, as in our
previous work \cite{Dar03}, the effective Lagrangian model developed
by Schmidt {\rm et al.}  \cite{ScA96}. The main advantage of this
model is that it has been constructed to give a realistic description
of the $\Delta$(1232)-resonance region. It is also given in an
arbitrary frame of reference and allows a well defined off-shell
continuation, as required for studying pion production reactions from
nuclei. This model consists of the standard pseudovector Born terms
and the contribution of the $\Delta(1232)$-resonance. For further
details with respect to the elementary pion photo-production operator,
we refer the reader to Ref.\ \cite{ScA96}. As shown in Figs.\ 1 - 3 of
our previous work \cite{Dar03}, the results of our calculations for
the elementary process are in good agreement with recent experimental
data, as well as with other theoretical predictions.  This gives a
clear indication that this elementary operator is quite satisfactory
for our purpose, namely to incorporate it into the pion
photo-production reaction from the deuteron.

For nucleon-nucleon scattering in the $NN$-subsystem, we use in this
work a specific class of separable potentials \cite{HaP8485} which
historically have played and still play a major role in the
development of few-body physics and also fit the phase shift data for
$NN$-scattering. The EST method \cite{Ern7374} for constructing
separable representations of modern $NN$ potentials has been applied
by the Graz group \cite{HaP8485} to cast the Paris potential
\cite{La+80} into a separable form. This separable model is most
widely used in the case of the $\pi NN$ system (see, for example,
Ref.\ \cite{Gar90} and references therein). Therefore, for the present
study of the influence of $NN$-rescattering, this model is
appropriate.
\section{$\pi$-photoproduction from the deuteron}\label{sec3}
The formalism of incoherent pion photo-production reaction from the
deuteron is presented in detail in our previous work \cite{Dar03}.
Here, we briefly recall the necessary notation and definitions. As
shown in Ref.\ \cite{BjD64}, the general expression for the
unpolarized cross section is given by
\begin{figure}[htb]
\includegraphics[scale=1.0]{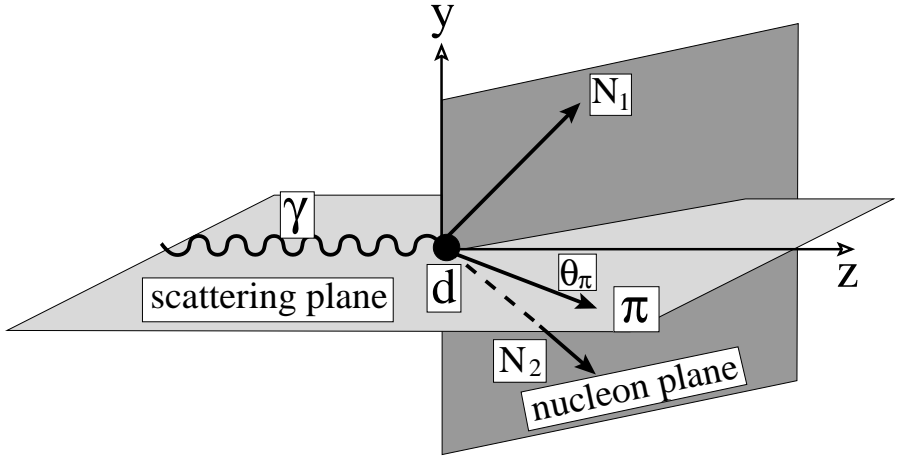}
  \caption{\small Kinematics in the laboratory system for pion
  photo-production from the deuteron.}
  \label{labsys}
\end{figure}
\begin{eqnarray}
d\sigma &=&
\frac{\delta^{4}(k+d-p_1-p_2-q)M^2_{N}d^{3} p_1d^{3} p_2d^{3} q}
{96(2 \pi)^{5} |\vec v_{\gamma} - \vec
v_{d}|\omega_{\gamma}E_{d}E_1E_2\omega_q} 
\hspace{-0.2cm}\sum_{s\,m\,t\,,m_{\gamma}\,m_d} 
\left |{\mathcal M}^{(t\,\mu)}_{s\,m\,m_{\gamma}\,m_d}
(\vec p_1,\vec p_2,\vec q,\vec k,\vec d) \right|^{2}\,,
\nonumber \\ & &
\label{eq:3.2}
\end{eqnarray}
where $k=(\omega_{\gamma},\vec k)$, $d=(E_d,\vec d)$,
$q=(\omega_q,\vec q)$, $p_1=(E_1,\vec p_1)$ and $p_2=(E_2,\vec p_2)$
denote the 4-momenta of the photon, deuteron, pion and two nucleons,
respectively. Furthermore, $m_{\gamma}$ denotes the photon
polarization, $m_{d}$ the spin projection of the deuteron, $s$ and $m$
the total spin and projection of the two outgoing nucleons, respectively,
$t$ their total isospin, $\mu$ the isospin projection of the pion, and
$\vec{v}_{\gamma}$ and $\vec{v}_{d}$ the velocities of the photon and
deuteron, respectively. The transition amplitude is denoted by
${\mathcal M}$. Covariant state normalization according to the 
convention of Ref.\ \cite{BjD64} is assumed. 

This expression is evaluated in the lab or deuteron rest frame.  A
right-handed coordinate system is chosen, where the $z$-axis is
defined by the photon momentum $\vec k$ and the $y$-axis by $\vec k
\times \vec q$. The scattering plane is defined by the momenta of
photon $\vec k$ and pion $\vec q$, whereas the momenta of the outgoing
nucleons $\vec p_1$ and $\vec p_2$ define the nucleon plane (see Fig.\ 
\ref{labsys}). As independent variables, the pion momentum $q$, its
angles $\theta_{\pi}$ and $\phi_{\pi}$, the polar angle
$\theta_{p_{NN}}$ and the azimuthal angle $\phi_{p_{NN}}$ of the
relative momentum $\vec p_{NN}$ of the two outgoing nucleons are
chosen. The total and relative momenta of the final $NN$-system are
defined by $\vec{P}_{NN} = \vec{p}_{1} + \vec{p}_{2}= \vec{k} -
\vec{q}$ and $\vec p_{NN} = \frac{1}{2}\left(\vec{p}_{1} -
  \vec{p}_{2}\right)$, respectively.

Integrating over the pion momentum $q$ and over $\Omega_{p_{NN}}$, one
obtains the semi-inclusive differential cross section of pion
photo-production from the deuteron, where only the final pion is detected
without analyzing its energy,  
\beq
\frac{d\sigma}{d\Omega_{\pi}}=\int_0^{q_{\rm max}}dq
\int d\Omega_{p_{NN}}\,
\frac{\rho_{s}}{6}\,\sum_{s\,m\,t\,m_{\gamma}\,m_{d}}  
\left \vert {\mathcal
M}^{(t\,\mu)}_{s\,m\,m_{\gamma}\,m_{d}}(\vec{p}_{NN},\vec q,\vec k)
\right \vert^{2}\,, 
\eeq
where $\rho_{s}$ denotes the phase space factor [see Eq.\ (7) in 
Ref.\ \cite{Dar03} for its definition].
\begin{figure}[htb]
\hspace*{1cm}\includegraphics[scale=.8]{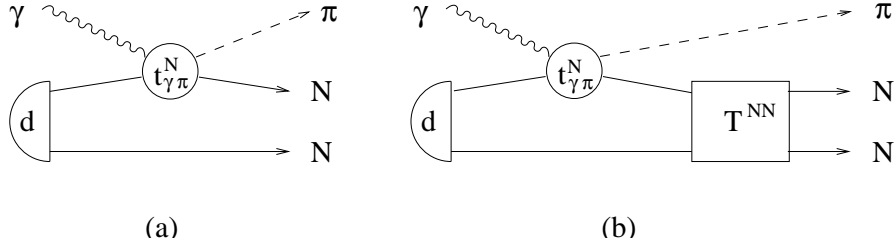}
\caption{Diagrammatic representation of pion photo-production from the
deuteron including $NN$-rescattering in the final state: (a) impulse
approximation (IA) and (b) $NN$-rescattering.} 
\label{t-matrix}
\end{figure}

The general form of the photo-production transition matrix is given by
\begin{eqnarray}\label{general}
{\mathcal M}^{(t\mu)}_{sm
m_{\gamma}m_d}(\vec{k},\vec{q},\vec{p_1},\vec{p_2}) 
& = &
^{(-)}\bra{\vec{q}\,\mu,\vec{p_1}\vec{p_2}\,s\,m\,t-\mu}\epsilon_{\mu}  
(m_{\gamma})J^{\mu}(0)\ket{\vec{d}\,m_d\,00}\,,
\end{eqnarray}
where $J^{\mu}(0)$ denotes the current operator. The outgoing $\pi NN$
scattering state is approximated in this 
work by  
\begin{eqnarray}
\ket{\vec{q}\,\mu,\vec{p_1}\vec{p_2}\,s\,m\,t-\mu}^{(-)} &=&
\ket{\vec{q}\,\mu,\vec{p_1}\vec{p_2}\,s\,m\,t-\mu} 
+ G_{0}^{\pi NN (-)}
\,T^{NN}
\ket{\vec{q}\,\mu,\vec{p_1}\vec{p_2}\,s\,m\,t-\mu}\,,\nonumber \\ & & 
\end{eqnarray}
where $\ket{\vec{q}\,\mu,\vec{p_1}\vec{p_2}\,s\,m\,t-\mu}$ denotes the 
free $\pi NN$ plane wave, $G_{0}^{\pi NN (-)}$ the free $\pi NN$
propagator, and $T^{NN}$ the reaction operator for $NN$-scattering. Thus,
the total transition matrix element in this approximation reads 
\begin{eqnarray}
\label{threethree}
{\mathcal M}^{(t\mu)}_{sm m_{\gamma}m_d} & = &
{\mathcal M}_{sm m_{\gamma}m_d}^{(t\mu)~IA} + 
{\mathcal M}_{sm m_{\gamma}m_d}^{(t\mu)~NN}\,. 
\end{eqnarray}
A graphical representation of the transition matrix is given in
Fig.~\ref{t-matrix}.

As shown in Ref.\ \cite{Dar03}, the matrix element in the IA is given by the 
expression  
\begin{eqnarray}\label{g16}
  {\mathcal M}_{sm m_{\gamma}m_d}^{(t\mu)~IA}
  (\vec k,\vec q,\vec p_1,\vec p_2) &=&
 \sqrt{2}\sum_{m^{\prime}}\langle s 
  m,\,t -\mu|\,\Big( \langle
  \vec{p}_{1}|t_{\gamma\pi}(\vec k,\vec q\,)|-\vec{p}_{2}\rangle
  \tilde{\Psi}_{m^{\prime},m_{d}}(\vec{p}_{2}) 
\nonumber\\ & &  \hspace{1cm} 
-(-)^{s+t}(\vec p_1 \leftrightarrow \vec p_2) 
\Big)\,|1
  m^{\prime},\,00\rangle \,,
\end{eqnarray}
where $t_{\gamma\pi}$ denotes the elementary production amplitude from 
the free nucleon and $\widetilde{\Psi}_{m,m_{d}}(\vec{p}\,)$ is given by 
\begin{equation}
  \widetilde{\Psi}_{m,m_{d}}(\vec{p}\,) =
  (2\pi)^{\frac{3}{2}}\sqrt{2E_{d}}
  \sum_{L=0,2}\sum_{m_{L}}i^{L}\,C^{L 1 1}_{m_{L} m m_{d}}\,
  u_{L}(p)Y_{Lm_{L}}(\hat{p}) \,.
\end{equation}
For the radial deuteron wave function $u_{L}(p)$, the Paris potential
\cite{La+81} is used.

For the $NN$-rescattering contribution, one obtains \cite{Dar03}
\begin{eqnarray}
\label{tnn-fsi-final}
{\mathcal M}^{(t\mu)~NN}_{sm m_{\gamma}m_d}
(\vec k,\vec q,\vec p_1,\vec p_2) & = & 
\sum_{m^{\prime}}\int d^3\vec p^{\,\prime}_{NN} 
\sqrt{\frac{E_1 E_2}{E_1' E_2'}}
\,\widetilde {\mathcal R}_{s m m^{\prime}}^{NN,\,t\mu}(W_{NN},\vec
p_{NN},\vec p^{\,\prime}_{NN}) \nonumber\\
&& \times~\frac{M_N}{\widetilde p^{\, 2} - p_{NN}^{\prime\,2} + i\epsilon}
{\mathcal M}^{(t\mu)~IA}_{sm^{\prime},m_{\gamma}m_d}(\vec k,\vec
q,\vec p^{\,\prime}_1,\vec p^{\,\prime}_2)\,,
\end{eqnarray}
where $\vec p_{NN}^{\,\prime}=\frac{1}{2}\,(\vec{p}_1^{\,\prime}-
\vec{p}_2^{\,\prime})$ denotes the relative momentum of the
interacting nucleons in the intermediate state, $W_{NN}$ is the
invariant mass of the $NN$-subsystem, $\vec
p^{\,\prime}_{1/2}=\pm \vec p^{\,\prime}_{NN} + (\vec k-\vec q\,)/2$
and $E_{1/2}'$ are the momenta and the corresponding on-shell energies
of the two nucleons in the intermediate state, respectively, and
$\widetilde p^{\,2} = M_N(E_{\gamma
d}-\omega_{\pi}-2M_N-(\vec{k}-\vec{q})^2/4M_N)$, with $E_{\gamma
d}=M_d+\omega_{\gamma}$. The conventional $NN$-scattering matrix
$\widetilde {\mathcal 
R}_{smm^{\prime}}^{NN,\,t\mu}$ is introduced with respect to
noncovariantly normalized states. It is expanded in terms of the
partial wave contributions ${\mathcal
T}_{Js\ell\ell^{\prime}}^{NN,\,t\mu}$  as  
\begin{eqnarray}
\label{tnn-hos}
\widetilde {\mathcal R}_{smm^{\prime}}^{NN,\,t\mu}(W_{NN},\vec
 p_{NN},\vec p^{\,\prime}_{NN}) & = & \sum_{J\ell\ell^{\prime}}
 {\mathcal F}_{\ell\ell^{\prime}\,mm'}^{NN,\,Js}
 (\hat{p}_{NN},\hat{p}_{NN}^{\,\prime}) \nonumber \\
 & & \times~{\mathcal T}_{Js\ell\ell^{\prime}}^{NN,\,t\mu}
 (W_{NN},p_{NN},p_{NN}^{\,\prime})\, , 
\end{eqnarray}
where the purely angular function ${\mathcal
F}_{\ell\ell^{\prime} mm^{\prime}}^{NN,\,Js}
(\hat{p}_{NN},\hat{p}_{NN}^{\,\prime})$ is defined by  
\begin{eqnarray}
\label{tnn-ff}
{\mathcal F}_{\ell\ell^{\prime} mm^{\prime}}^{NN,\,Js}(\hat{p}_{NN},
\hat{p}_{NN}^{\,\prime}) & = & 
 \sum_{M m_{\ell}m_{\ell^{\prime}}} C^{\ell s J}_{m_{\ell} m M}\,
C^{\ell^{\prime} s J}_{m_{\ell^{\prime}} m^{\prime} M}
 Y^{\star}_{\ell m_{\ell}}(\hat{p}_{NN})
 Y_{\ell^{\prime} m_{\ell^{\prime}}}(\hat{p}_{NN}^{\,\prime})\,.
\end{eqnarray}
The necessary half-off-shell $NN$-scattering matrix ${\mathcal
T}_{Js\ell\ell^{\prime}}^{NN,\,t\mu}$ was obtained from the separable
representation of a realistic $NN$-interaction \cite{HaP8485} which
gives a good description of the corresponding phase shifts. Explicitly,
all partial waves with  total angular momentum $J\le 3$ have been
included. 
\section{Results and discussion}\label{sec5}
The discussion of our results is divided into two parts. First, we
discuss the influence of the $NN$-FSI effect on the polarized
differential cross-section difference
$({d\sigma/d\Omega_{\pi}})^P-({d\sigma/d\Omega_{\pi}})^A$ for the
parallel and antiparallel helicity states by comparing the pure IA
with the inclusion of $NN$-rescattering in the final state.
Furthermore, we compare our results with recent experimental data from
the GDH collaboration \cite{Pedroni}. In the second part, we consider
the polarized total cross sections for circularly polarized photons on
a target whose spin is parallel $\sigma^P$ and antiparallel $\sigma^A$
to the photon spin. The contribution of $\vec\gamma\vec d\to\pi^-pp$
to the spin response of the deuteron, i.e., the asymmetry of the total
photo-absorption cross section with respect to parallel and
antiparallel spins of photon and deuteron, has been explicitly
evaluated over the range of the $\Delta$(1232)-resonance with
inclusion of final-state $NN$-rescattering.
\subsection{The helicity difference
$({d\sigma/d\Omega_{\pi}})^P-({d\sigma/d\Omega_{\pi}})^A$}\label{sec51} 
\begin{figure}[htp]
\begin{center}
\includegraphics[scale=0.8]{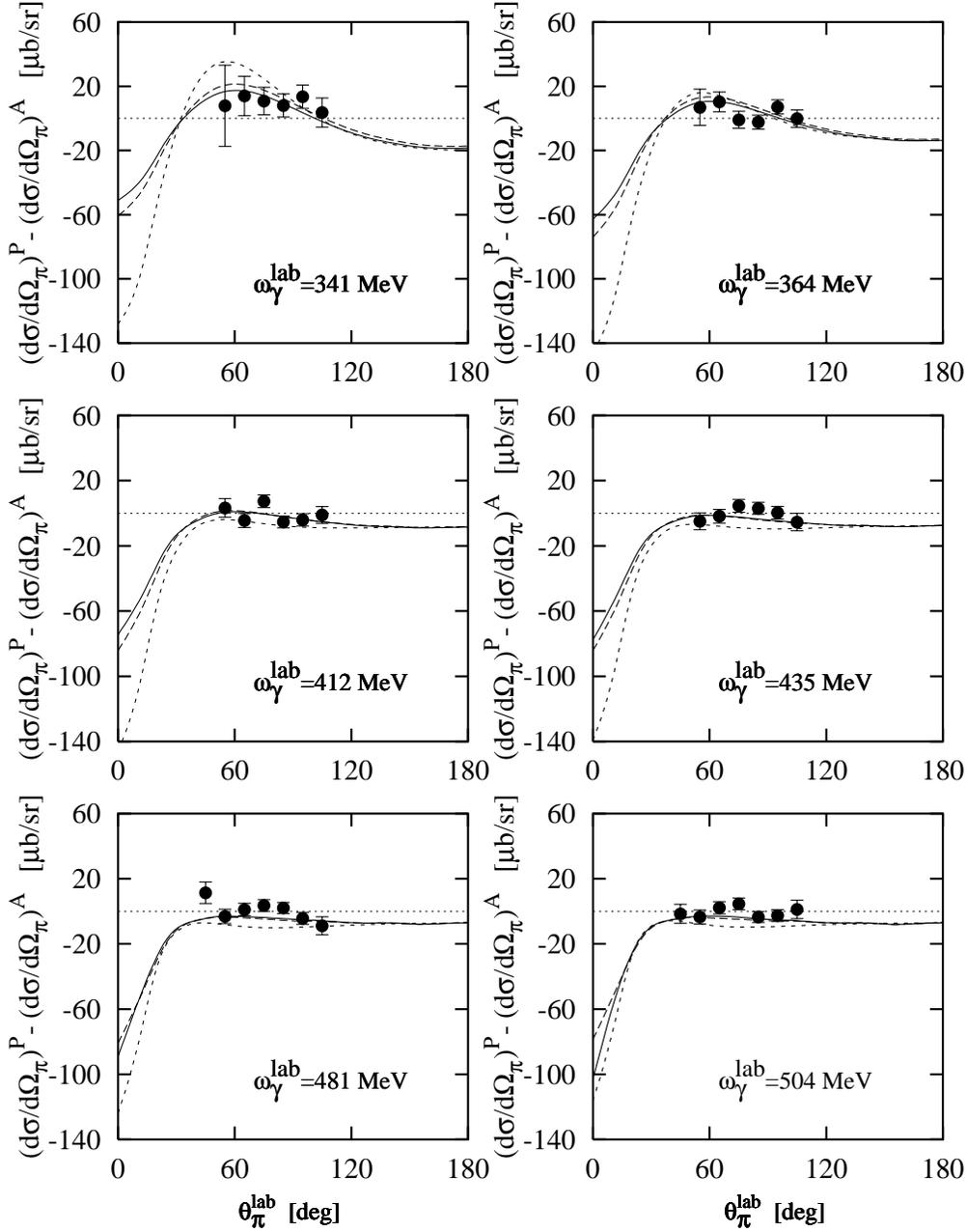}
\caption{The differential polarized cross-section difference
$({d\sigma/d\Omega_{\pi}})^P-({d\sigma/d\Omega_{\pi}})^A$ for
$\vec\gamma\vec d\to\pi^-pp$ for the parallel 
$({d\sigma/d\Omega_{\pi}})^P$ and antiparallel
$({d\sigma/d\Omega_{\pi}})^A$ helicity states as a function of
the pion angle in the laboratory frame in comparison with recent measurements 
presented in Ref.\ \cite{Pedroni} at different values of the photon
lab energy. Dashed curves: IA; solid
curves: IA+$NN$-rescattering; dotted curves: predictions for $\pi^-$
production from the free neutron, i.e., $\vec\gamma\vec n\to\pi^-p$.} 
\label{diff}
\end{center}
\end{figure}

\begin{figure}[htp]
\begin{center}
\includegraphics[scale=0.8]{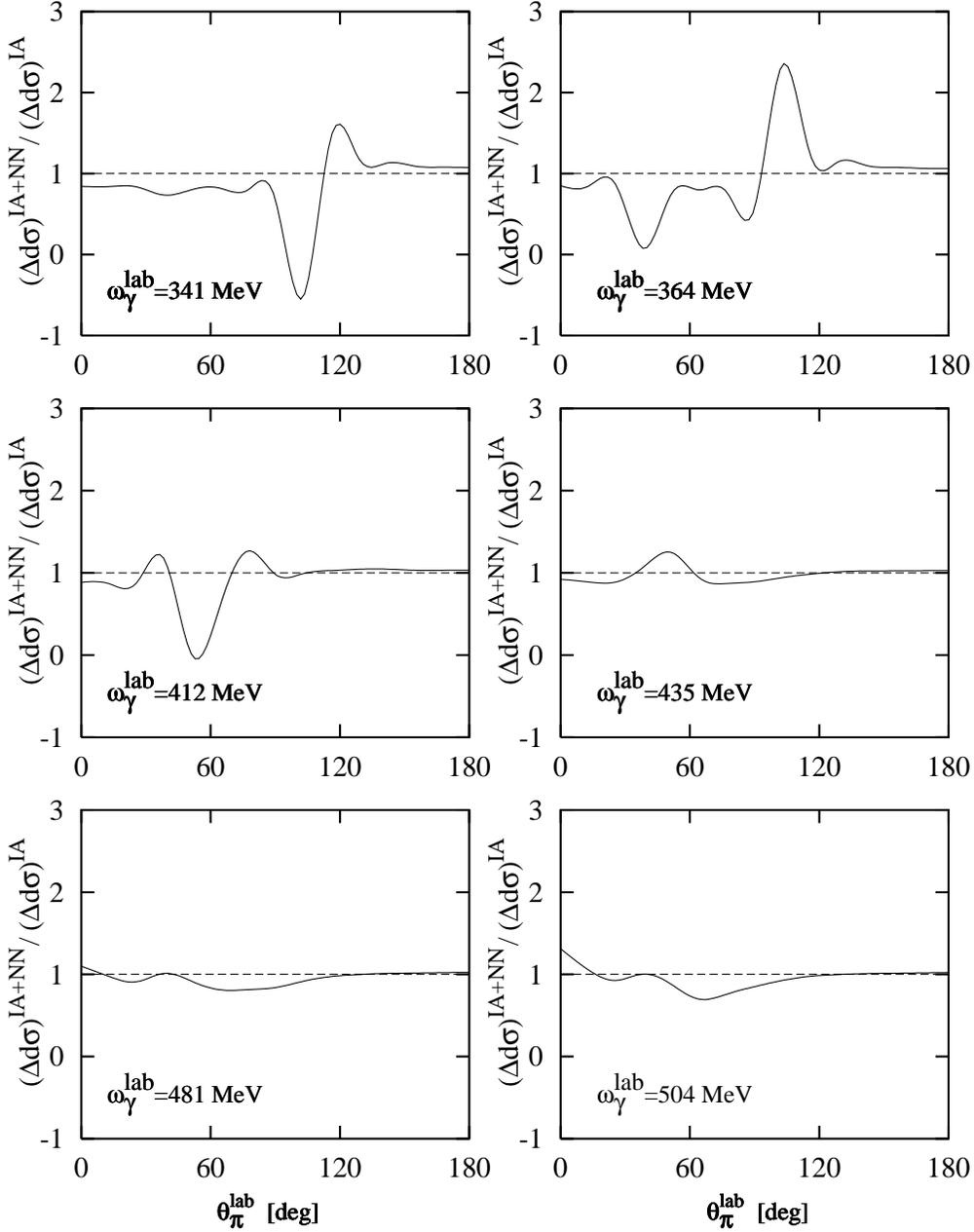}
\caption{The ratio $(\Delta d\sigma)^{IA+NN}/(\Delta d\sigma)^{IA}$ 
[see Eq.\ (\ref{ratio}) for its definition] as a function of the pion
angle in the laboratory frame for several photon lab energies.} 
\label{diffdiff}
\end{center}
\end{figure}

We begin the discussion by presenting our results for the differential
polarized cross-section difference for parallel
$({d\sigma/d\Omega_{\pi}})^P$ and antiparallel
$({d\sigma/d\Omega_{\pi}})^A$ helicity states in pure IA and with
$NN$-rescattering, as shown in Fig.\ \ref{diff} as a function of the
emission pion angle in the laboratory frame at several values of the
photon lab energy. It is readily seen that $NN$-rescattering---the
difference between the dashed and the solid curves---is quite small,
and indeed almost completely negligible at pion backward angles. The
reason for this stems from the fact that in charged-pion production,
$^3S_1$-contribution to the $NN$ final state is forbidden.
In order to make a more detailed and quantitative evaluation of
$NN$-FSI with respect to the differential polarized cross-section
difference, we display in Fig.\ \ref{diffdiff} the relative effect by
plotting the ratio of the corresponding cross-section difference to
those for the IA, i.e., 
\begin{eqnarray}
\frac{(\Delta d\sigma)^{IA+NN}}{\hspace*{-0.6cm}(\Delta d\sigma)^{IA}}
&~=~&  
\frac{\left[(\frac{d\sigma}{d\Omega_{\pi}})^P -
(\frac{d\sigma}{d\Omega_{\pi}})^A\right]^{IA+NN}}
{\hspace*{-0.6cm}\left[(\frac{d\sigma}{d\Omega_{\pi}})^P -
(\frac{d\sigma}{d\Omega_{\pi}})^A\right]^{IA} } \,.
\label{ratio}
\end{eqnarray}
It is seen that the major contribution
from $NN$-FSI appears at forward pion angles. This
contribution is much less important in the differential polarized
cross-section difference than in the previously studied unpolarized
differential cross sections. (Compare with Fig.\ 13 in Ref.\ \cite{Dar03}.) It has been found that $NN$-FSI reduces the unpolarized 
differential cross section by about 15$\%$ at
$\theta_{\pi}=0^{\circ}$ \cite{Dar03}. The magnitude of this reduction decreases
rapidly with increasing pion angle. 

By comparing the results for the difference
$({d\sigma/d\Omega_{\pi}})^P-({d\sigma/d\Omega_{\pi}})^A$ in the case
of $\vec\gamma\vec d\to\pi^-pp$ (solid curves in Fig.\ \ref{diff})
with those in the case of the free $\vec\gamma\vec n\to\pi^-p$ (dotted
curves in Fig.\ \ref{diff}), we see that a large correction is needed
to go from the bound deuteron to the free neutron case. The difference
between the two results decreases to a tiny effect at backward angles.
Figure \ref{diff} also gives a comparison of our results for the
helicity difference with the experimental data from the GDH
collaboration \cite{Pedroni}. It is obvious that quite satisfactory
agreement with experiment is achieved. An experimental check of the
helicity difference at extreme forward and backward pion angles is
needed. Also, an independent check in the framework of effective field
theory would be very interesting.
\subsection{Polarized total cross sections}\label{sec52}
\begin{figure}[htp]
\begin{center}
\includegraphics[scale=0.8]{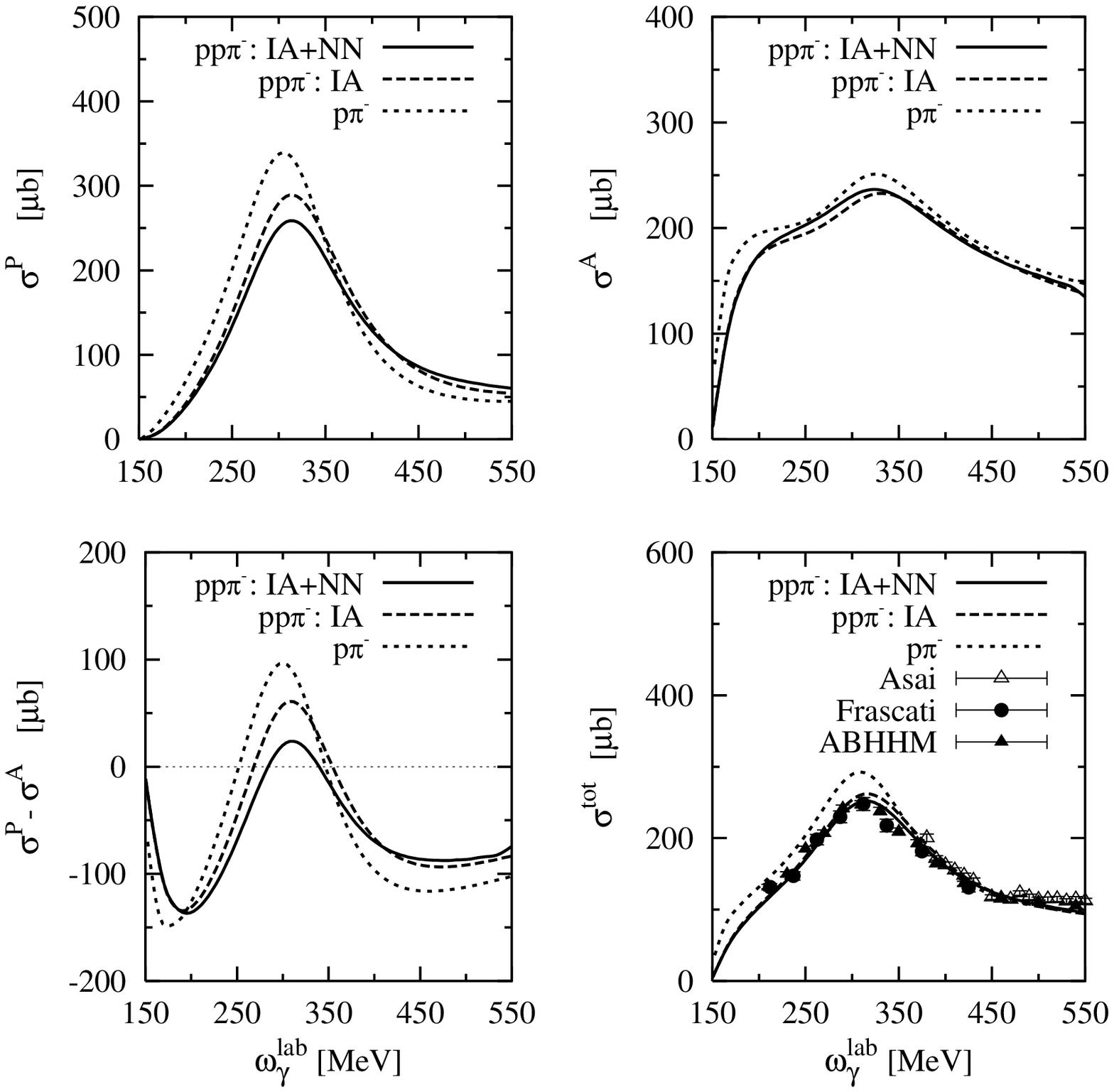}
\caption{The total photo-absorption cross sections for
circularly polarized photons impinging on a target with spin parallel $\sigma^P$ 
(upper left part) and antiparallel $\sigma^A$ (upper right part) to
the photon spin for $\vec\gamma \vec d\to\pi^-pp$ as 
functions of the photon lab energy. The lower part displays the difference
$\sigma^P-\sigma^A$ (lower left part) and the unpolarized total cross
section (lower right part). The experimental data are 
from Refs.\ \cite{Be+73} (ABHHM), \cite{ChD75} (Frascati) and \cite{As+90}
(Asai). The identification of the curves is the same as in Fig.\ \ref{diff}.}  
\label{total}
\end{center}
\end{figure}

\begin{figure}[htp]
\begin{center}
\includegraphics[scale=0.8]{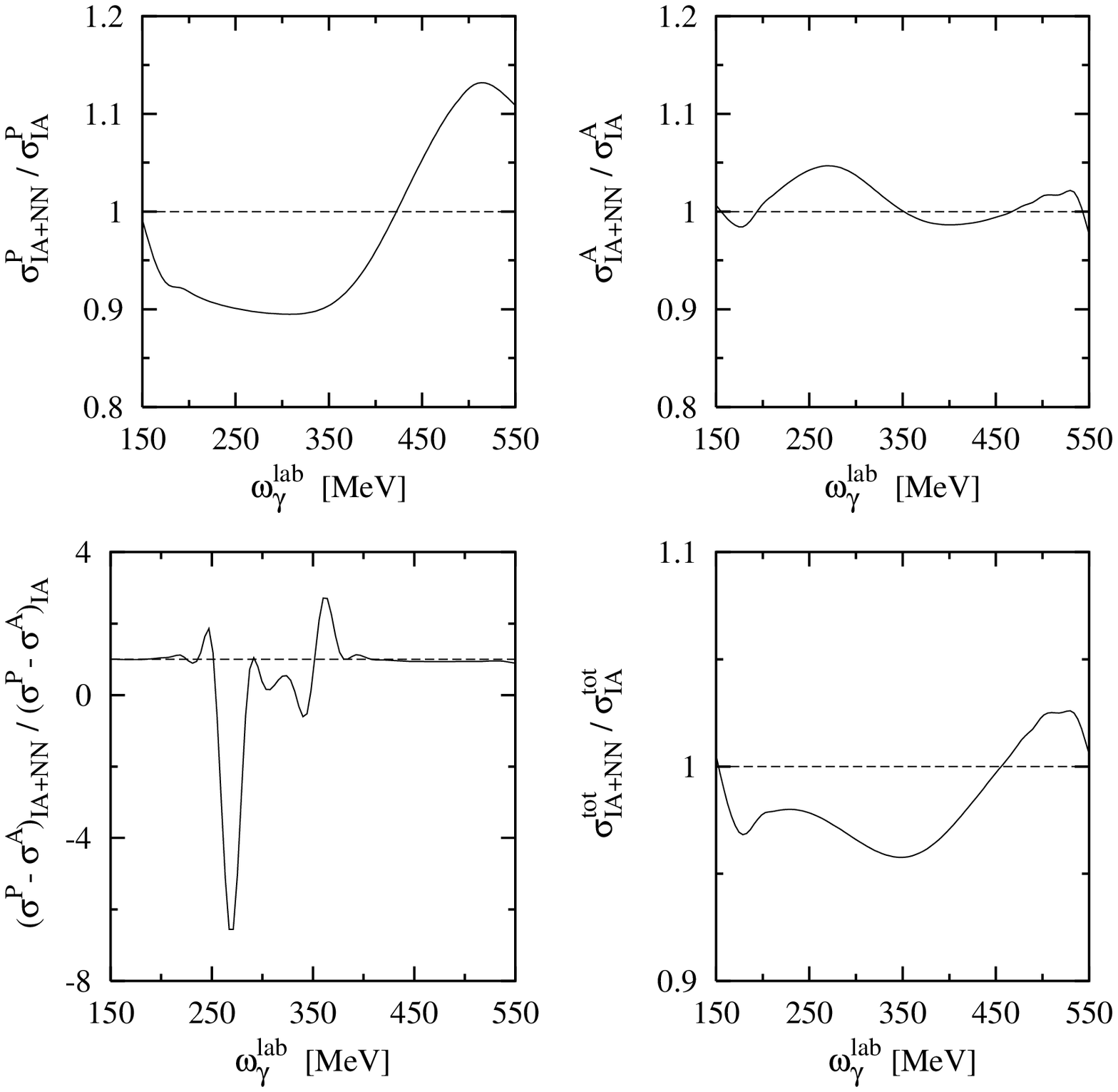}
\caption{The ratios
$\sigma^P_{IA+NN}/\sigma^P_{IA}$ (upper left part),  
$\sigma^A_{IA+NN}/\sigma^A_{IA}$ (upper right part),  
$(\sigma^P-\sigma^A)_{IA+NN}/(\sigma^P-\sigma^A)_{IA}$ (lower left part) 
and $\sigma^{tot}_{IA+NN}/\sigma^{tot}_{IA}$ (lower right part)
as functions of the photon lab energy.} 
\label{totalratio}
\end{center}
\end{figure}

Here, we discuss the results for the polarized total cross sections in
the case of the IA alone and with the $NN$-FSI effect. They are
presented in Fig.\ \ref{total}, where the left top panel displays the
total photo-absorption cross section $\sigma^P$ for circularly
polarized photons impinging on a target with spin parallel to the
photon spin, the right top panel displays that for antiparallel spins
of photon and target $\sigma^A$, the left bottom panel displays the
spin asymmetry $\sigma^P-\sigma^A$, and the right bottom panel
displays the results for the unpolarized total cross section in
comparison with the experimental data from Refs.\ \cite{Be+73}
(ABHHM), \cite{ChD75} (Frascati) and \cite{As+90} (Asai). For
comparison, we also depict in the same figure the results for $\pi^-$
production from the free neutron target by the dotted curves.
In order to see more clearly the relative size of the
interaction effect, we have plotted in Fig.\ \ref{totalratio} the
ratios with respect to the IA.

We find for the cross sections $\sigma^P$ and $\sigma^A$, the spin
asymmetry $\sigma^P-\sigma^A$, and for the unpolarized total cross
section of the nucleon and the deuteron qualitatively similar
behaviour, although for the deuteron, the maxima and minima are
smaller and also slightly shifted toward higher energies. Furthermore,
in the case of $\sigma^P$, a large deviation between the results for
the IA and the elementary reaction---the difference between the dashed
and the dotted curves---is seen because of the Fermi motion and FSI,
whereas for $\sigma^A$ the difference is smaller. The $NN$-FSI effect
appears mainly in $\sigma^P$.
The left bottom panel in Fig.\ \ref{total} shows that the helicity
difference of the total cross section ($\sigma^P-\sigma^A$) starts out
negative due to the $E_{0+}$ multipole, which is dominant in the
threshold region and has a strong positive contribution due to the
$M_{1+}$ multipole, which is dominant in the $\Delta$(1232)-resonance
region. It is also clear that FSI leads to a strong reduction of the
spin asymmetry in the energy region of the $\Delta$(1232)-resonance.
This reduction becomes about 35 $\mu$b at its maximum. Thus, the IA is
not a reasonable approximation, as in the case of the unpolarized
total cross section. Moreover, already the IA deviates significantly
from the corresponding nucleon quantities. It is also obvious that
$\sigma^P$ is much larger than $\sigma^A$ because of the
$\Delta$-excitation.

For the unpolarized total cross section displayed in the bottom right
panel of Fig.\ \ref{total}, it is also seen that the $NN$-FSI effect
is small, not more than about 5 percent. This effect comes mainly from
the change in the radial wave function of the final $NN$ partial waves
caused by the interaction. Therefore, it reduces the cross section.
The charged final state $p\pi^-$ was investigated 30 years ago in a
bubble chamber experiment on the $\gamma d\to pp\pi^-$ reaction by the
ABHHM collaboration \cite{Be+73}, at Frascatti \cite{ChD75}, and later
at higher energies by the TAGX-collaboration \cite{As+90}.  The bottom
right panel of Fig.\ \ref{total} presents a comparison between our
results and this set of experimental data. It is readily seen that the
inclusion of $NN$-rescattering considerably improves the agreement
between the experimental data and the theoretical predictions.
\section{Summary and conclusions}\label{sec6}
We have investigated the influence of the $NN$-FSI effect on the
polarized differential and total cross-section differences
$({d\sigma/d\Omega_{\pi}})^P-({d\sigma/d\Omega_{\pi}})^A$ and
$\sigma^P-\sigma^A$, respectively, for parallel and antiparallel
helicity states for the $\vec\gamma\vec d\to\pi^- pp$ reaction.  These
helicity asymmetries give valuable information concerning the nucleon
spin structure and allow a test of the GDH sum rule. For the
elementary pion production operator from the free nucleon, we used an
effective Lagrangian model. As the model for the interaction of the
$NN$-subsystem, we used a separable representation of a realistic $NN$
interaction, which gives a good description of the corresponding phase
shifts.

The study of the polarized differential cross-section difference
reveals that the reduction realized by including the $NN$-rescattering
is 15$\%$ larger at pion forward angles. For pions emitted in the
backward direction, the $NN$-rescattering effect is completely
negligible. In comparison with experiment, quite satisfactory
agreement is obtained.  The polarized total cross sections for
circularly polarized photons impinging on a target with spin parallel
$\sigma^P$ and antiparallel $\sigma^A$ to the photon spin are also
investigated. The contribution of $\vec\gamma\vec d\to\pi^-pp$ to the
spin response of the deuteron has been explicitly evaluated over the
range of the $\Delta$(1232)-resonance with inclusion of
$NN$-rescattering.  In the case of $\sigma^P$, we obtained a
significant difference between the results for the IA and the
elementary reaction, whereas for $\sigma^A$ the difference is smaller.
We found that $NN$-FSI effect appears mainly in $\sigma^P$.  It leads
to a strong reduction of the spin asymmetry in the energy region of
the $\Delta$(1232)-resonance. This reduction becomes about 35 $\mu$b
at its maximum. For the unpolarized total cross section, we found that
$NN$-rescattering reduces the total cross section in the
$\Delta$(1232)-resonance region by about 5 percent. In comparison with
experiment, the inclusion of such an effect leads to improved
agreement with the experimental data.

It remains as a task for further theoretical research to investigate
the reaction $\gamma d\to\pi NN$ including a three-body treatment in
the final $\pi NN$ system. This extension is desirable for the
calculation of such rescattering to help in further developments.
Instead of a separable potential, a more realistic potential for the
$NN$-scattering should be considered. A further interesting topic
concerns the study of polarization observables with the inclusion of
rescattering effects. Such studies should give more detailed
information on the $\pi NN$ dynamics and thus provide more stringent
tests for theoretical models. As future refinements, we consider also
the use of a more sophisticated elementary production operator, which
will allow for the extension of the present results to higher
energies. A measurement of the spin asymmetry for the deuteron is
needed.
\section*{Acknowledgements}
I gratefully acknowledge very useful discussions with H.\
Arenh\"ovel as well as the members of his group. I would like to  
thank the members of the GDH collaboration, especially P.\ Pedroni, 
for providing us with their experimental data.

\end{document}